\documentclass{article}

\usepackage{microtype}
\usepackage{graphicx}
\usepackage{subfigure}
\usepackage{booktabs} 

\usepackage[breaklinks=true]{hyperref}

\usepackage[arxiv]{icml2024}

\usepackage{amsmath}
\usepackage{amssymb}
\usepackage{mathtools}
\usepackage{amsthm}

\usepackage[capitalize,noabbrev]{cleveref}

\theoremstyle{plain}

\theoremstyle{definition}

\theoremstyle{remark}

\usepackage[textsize=tiny]{todonotes}

\begin{document}

\twocolumn[
\icmltitle{FLUX that Plays Music}



\icmlsetsymbol{equal}{*}

\begin{icmlauthorlist}
\icmlauthor{Zhengcong Fei}{}\;
\icmlauthor{Mingyuan Fan}{}\;
\icmlauthor{Changqian Yu}{}\;
\icmlauthor{Junshi Huang}{sch} \\
Kunlun Inc.
\end{icmlauthorlist}

\icmlaffiliation{sch}{Corresponding author}
\icmlcorrespondingauthor{Junshi Huang}{feizhengcong@gmail.com}
\icmlkeywords{Machine Learning, ICML}

\vskip 0.3in
]




\begin{abstract}

This paper explores a simple extension of diffusion-based rectified flow Transformers for text-to-music generation, termed as FluxMusic.  
Generally, along with design in advanced Flux\footnote{https://github.com/black-forest-labs/flux} model, we transfers it into a latent VAE space of mel-spectrum. It involves first applying a sequence of independent attention to the double text-music stream, followed by a stacked single music stream for denoised patch prediction. We employ multiple pre-trained text encoders to sufficiently capture caption semantic information as well as inference flexibility. In between, coarse textual information, in conjunction with time step embeddings, is utilized in a modulation mechanism, while fine-grained textual details are concatenated with the music patch sequence as inputs. 
Through an in-depth study, we demonstrate that rectified flow training with an optimized architecture significantly outperforms established diffusion methods for the text-to-music task, as evidenced by various automatic metrics and human preference evaluations. 
Our experimental data, code, and model weights are made publicly available at: \url{https://github.com/feizc/FluxMusic}.

\end{abstract}

\section{Introduction}

Music, as a form of artistic expression, holds profound cultural importance and resonates deeply with human experiences \cite{briot2017deep}. The task of text-to-music generation, which involves converting textual descriptions of emotions, styles, instruments, and other musical elements into audio, offers innovative tools and new avenues for multimedia creation \cite{huang2023noise2music}. Recent advancements in generative models have led to significant progress in this area \cite{yang2017midinet,dong2018musegan,mittal2021symbolic}. Traditionally, approaches to text-to-music generation have relied on either language models or diffusion models to represent quantized waveforms or spectral features \cite{agostinelli2023musiclm,lam2024efficient,liu2024audioldm,evans2024stable,schneider2024mousai,fei2024music,fei2023masked,chen2024musicldm}. Among these, diffusion models~\citep{Song2020ScoreBasedGM}, trained to reverse the process of data transformation from structured states to random noise~\citep{SohlDickstein2015DeepUL,song2020generative}, have shown exceptional effectiveness in modeling high-dimensional perceptual data, including music~\citep{ho2020denoising,huang2023make,ho2022video,kong2020diffwave,rombach2022high}.

Given the iterative nature of diffusion process, coupled with the significant computational costs and extended sampling times during inference, there has been a growing body of research focused on developing more efficient training strategy and accelerating sampling schedule ~\citep{karras2023analyzing,liu2022flow,lu2022dpm,fei2019fast,lu2022dpm2,kingma2024understanding}, such as distillation \cite{sauer2024fast,song2023consistency,song2023improved}. A particularly effective approach involves defining a forward path from data to noise, which facilitates more efficient training \cite{ma2024sit} as well as better generative performance. One effective method among them is the Rectified Flow \citep{liu2022flow,albergo2022building,lipman2023flow}, where data and noise are connected along a linear trajectory. It offers improved theoretical properties and has shown promising results in image generation\cite{ma2024sit,esser2024scaling}, however, its application in music creation remains largely unexplored.

In the design of model architectures, traditional diffusion models frequently employ U-Net \cite{ronneberger2015u} as the foundational structure. However, the inherent inductive biases of convolutional neural networks inadequately captures the spatial correlations within signals \cite{esser2021taming} and are insensitive to scaling laws~\cite{li2024scalability}. Transformer-based diffusion models have effectively addressed these limitations~\cite{Peebles_2023,bao2023all,fei2024scalable,fei2024diffusion} by treating images as sequences of concatenated patches and utilizing stacked transformer blocks for noise prediction. The incorporation of cross-attention  for integrating textual information has established this approach as the standard for generating high-resolution images and videos from natural language descriptions, demonstrating impressive generalization capabilities~\cite{chen2023pixart,chen2024pixarts,fei2024dimba,esser2024scaling,fei2023jepa,ma2024latte,ma2024latte,yang2024cogvideox}. Notably, the recently open-sourced FLUX model, with its well-designed structure, exhibits strong semantic understanding and produces high-quality images, positioning it as a promising framework for conditional generation tasks.

\begin{figure}[t]
  \centering
   \includegraphics[width=1\linewidth]{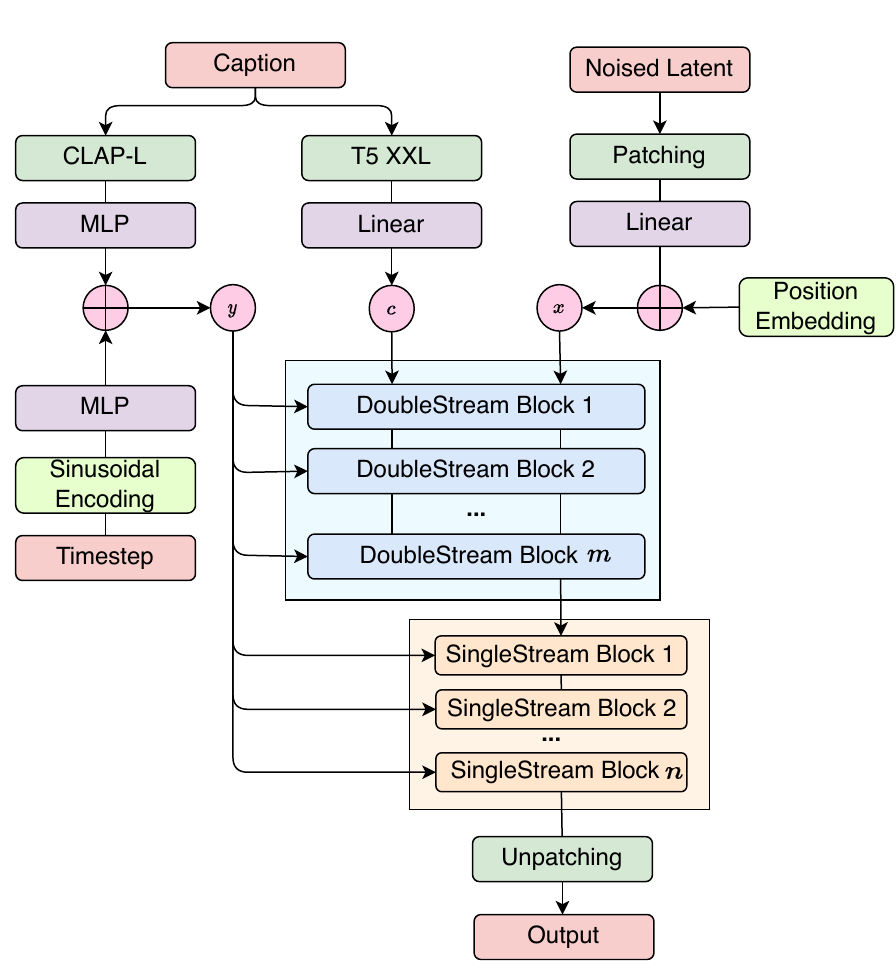}
   \caption{\textbf{Model architecture of FluxMusic. }  We use frozen CLAP-L and T5-XXL as text encoders for conditioned caption feature extraction. The coarse text information concatenated with timestep embedding $y$ are used to modulation mechanism. The fine-grained text $c$ concatenated with music sequence $x$ are input to a stacked of double stream block and single steam blocks to predict nose in a latent VAE space.
   }
   \label{fig:framework} 
\end{figure}

In this work, we explore the application of rectified flow Transformers within noise-predictive diffusion for text-to-music generation, introducing FluxMusic as a unified and scalable generative framework in the latent VAE space of the mel-spectrogram, as illustrated in Figure \ref{fig:framework}. 
Building upon the text-to-image FLUX model, we present a transformer-based architecture that initially integrates learnable double streams attention for the concatenated music-text sequence, facilitating a bidirectional flow of information between the modalities. Subsequently, the text stream is dropped, leaving a stacked single music stream for noised patch prediction. We leverage multiple pre-trained text encoders for extracting conditioned caption features and inference flexibility. Coarse textual information from CLAP-L~\cite{elizalde2023clap}, combined with time step embeddings, is employed in the modulation mechanism, while fine-grained textual details from T5-XXL~\cite{raffel2020exploring} are concatenated with music patch sequence as input. We train the model with rectified flow formulation and investigate its scalability. Through a in-depth study, we compare our new formulation to existing diffusion formulations and demonstrate its benefits for training efficiency and performance enhancement.

The primary contributions of this work are as follows: 
\begin{itemize}
    \item We introduce a Flux-like Transformer architecture for text-to-music generation, equipped with rectified flow training. To the best of our knowledge, this is the first study to apply rectified flow transformers to text-to-music generation;
    \item  We perform a comprehensive system analysis, encompassing network design, rectified flow sampling, and parameter scaling, demonstrating the advantages of FluxMusic architecture in text-to-music generation; 
    \item Extensive experimental results demonstrate that FluxMusic achieves generative performance on par with other recent  models with adequate training on both automatic metrics and human preference ratings. Finally, we make the results, code, and model weights publicly available to support further research.  
\end{itemize}

\section{Related Works}

\subsection{Text-to-music Generation}
Text-to-music generation seeks to produce music clips that correspond to descriptive or summarized text inputs. Prior approaches have primarily employed language models (LMs) or diffusion models (DMs) to generate quantized waveform representations or spectral features. For generating discrete representation of waveform, models such as MusicLM~\cite{agostinelli2023musiclm}, MusicGen~\cite{copet2024simple}, MeLoDy~\cite{lam2024efficient}, and JEN-1~\cite{li2024jen} utilize LMs and DMs on residual codebooks derived from quantization-based audio codecs~\cite{zeghidour2021soundstream,defossez2022high}. Conversely, models like Moûsai~\cite{schneider2024mousai}, Noise2Music~\cite{huang2023noise2music}, Riffusion~\cite{forsgren2022riffusion}, AudioLDM 2~\cite{liu2024audioldm}, MusicLDM~\cite{chen2024musicldm}, and StableAudio~\cite{evans2024stable} employ U-Net-based diffusion techniques to model mel-spectrograms or latent representations obtained through pretrained VAEs, subsequently converting them into audio waveforms using pretrained vocoders~\cite{kong2020hifi}. Additionally, models such as Mustango~\cite{melechovsky2023mustango} and Music Controlnet~\cite{wu2024music} incorporate control signals or personalization~\cite{plitsis2024investigating,fei2023gradient}, including chords and beats, in a manner similar to ControlNet \cite{zhang2023adding}. Our method along with this approach by modeling the mel-spectrogram within a latent VAE space.

\subsection{Diffusion Transformers}

Transformer architecture~\cite{vaswani2017attention} has achieved remarkable success in language models~\cite{radford2018improving,radford2019language,raffel2020exploring} and has also demonstrated significant potential across various computer vision tasks, including image classification~\cite{dosovitskiy2020image,he2022masked,touvron2021training,zhou2021deepvit,yuan2021tokens,han2021transformer}, object detection~\citep{liu2021swin,wang2021pyramid,wang2022pvt, carion2020end}, and semantic segmentation~\citep{zheng2021rethinking,xie2021segformer,strudel2021segmenter}, among others~\citep{sun2020transtrack,li2022panoptic,zhao2021point,liu2022video,he2022masked,li2022bevformer}. Building on this success, the diffusion Transformer~\cite{Peebles_2023,fei2024moe} and its variants~\cite{bao2023all,fei2024scalable} have replaced the convolutional-based U-Net backbone~\cite{ronneberger2015u} with Transformers, resulting in greater scalability and more straightforward parameter expansion compared to U-Net diffusion models. This scalability advantage has been particularly evident in domains such as video generation~\cite{ma2024latte}, image generation~\cite{chen2023pixart}, and speech generation~\cite{liu2023vit}. Notably, recent works such as Make-an-audio 2~\cite{huang2023make,huang2023make2} and StableAudio 2~\cite{evans2024stable} also explored the DiT architecture for audio and sound generation. In contrast, our work investigates the effectiveness of new multi-modal diffusion Transformer structure similar to Flux and optimized it with rectified flow.

\section{Methodology}

FluxMusic is a conceptually simple extension of FLUX, designed to facilitate text-to-music generation within a latent space. An overview of the model structure is illustrated in Figure \ref{fig:framework}. In the following, we begin with a review of rectified flow as applied to diffusion models, followed by a detailed examination of the architecture for each component. We also discuss considerations regarding model scaling and data quality.

\subsection{Rectified Flow Trajectories}

In this work, we explore generative models that estabilish a mapping between samples $x_1$ from a noise distribution $p_1$ to samples $x_0$ from a data distribution $p_0$ through a framework of ordinary differential equation (ODE). 
The connection can be expressed as $dy_t = v_\Theta(y_t, t)\,dt $ where the velocity $v$ is parameterized by the weights $\Theta$ of a neural network. 
\cite{Chen2018NeuralOD} proposed directly solving it using differentiable ODE solvers. However, it proves to be computationally intensive, particularly when applied to large neural network architectures that parameterize $v_\Theta(y_t, t)$.

A more effective strategy involves directly regressing a vector field $u_t$ that defines a probability trajectory between $p_0$ and $p_1$.
To construct such a vector field $u_t$, we consider a forward process that corresponds to a probability path $p_t$ transitioning from $p_0$ to $p_1=\mathcal{N}(0, 1)$. This can be represented as $z_t = a_t x_0 + b_t \epsilon\quad\text{, where}\;\epsilon \sim \mathcal{N}(0,I)$.
With the conditions $a_0 = 1, b_0 = 0, a_1 = 0$ and $b_1 = 1$, the marginals $p_t(z_t) =
  \mathbb{E}_{\epsilon \sim \mathcal{N}(0,I)}
  p_t(z_t \vert \epsilon)\;$ align with both the data and noise distribution.

Referring to \cite{lipman2023flow,esser2024scaling}, we can construct a marginal vector field $u_t$ that generates the marginal probability paths $p_t$, using the conditional vector fields as follows:
\begin{align}
    u_t(z) = \mathbb{E}_{\epsilon \sim
  \mathcal{N}(0,I)} [u_t(z \vert \epsilon) \frac{p_t(z \vert
  \epsilon)}{p_t(z)}],
  \label{eq:marginal_u}
\end{align}
The conditional flow matching objective can be then formulated as:
\begin{align}
   {L} =  \mathbb{E}_{t, p_t(z | \epsilon), p(\epsilon) }|| v_{\Theta}(z, t) - u_t(z | \epsilon)  ||_2^2\;, \label{eq:condflowmatch}
\end{align}
where the conditional vector fields $u_t(z \vert \epsilon)$ provides a tractable and equivalent objective. 
Although there exists different variants of the above formalism, we focus on Rectified Flows (RF) \cite{liu2022flow,albergo2022building,lipman2023flow}, which define the forward process as straight paths between the data distribution and a standard normal distribution: 
\begin{equation}
z_t = (1-t) x_0 + t \epsilon,
\end{equation}
with the loss function ${L}$ corresponds to $w_t^\text{RF} = \frac{t}{1-t}$. The network output directly parameterizes the velocity $v_\Theta$.

\begin{table*}[t]
\centering
\setlength{\tabcolsep}{2.mm}{
\begin{tabular}{lcccccc}
\toprule
& \#Params & \#DoubleStream $m$ &\#SingleStream $n$  & Hidden dim. $d$ & Head number $h$ &Gflops  \\ \midrule
Small  & 142.3M &8 & 16 & 512 & 16& 194.5G  \\
Base &473.9M &12 &24 &768 & 16 & 654.4G \\
Large &840.6M&12 &24 &1024 &16 & 1162.6G \\
Giant &2109.9M&16 &32 &1408 & 16  &2928.0G
\\ \bottomrule 
\end{tabular}}
\caption{\textbf{Scaling law of FluxMusic model size.} The model sizes and detailed hyperparameters settings for scaling experiments. 
}
\label{tab:scale}
\end{table*}

\subsection{Model Architecture}

To enable text-conditioned music generation, our FluxMusic model integrate both textual and musical modalities. We leverage pre-trained models to derive appropriate representations and then describe the architecture of our Flux-based model in detail.

\paragraph{Music compression.}
To better represent music, following \cite{liu2024audioldm}, each 10.24-second audio clip, sampled at 16kHz, is first converted into a $64 \times 1024$ mel-spectrogram, with 64 mel-bins, a hop length of 160, and a window length of 1024. This spectrogram is then compressed into a $16 \times 128$ latent representation, denoted as $X_{spec}$, using a Variational Autoencoder (VAE) pretrained on AudioLDM 2\footnote{https://huggingface.co/cvssp/audioldm2-music/tree/main}. This latent space representation serves as the basis for noise addition and model training. Finally, a pretrained Hifi-GAN \cite{kong2020hifi} is employed to reconstruct the waveform from the generated mel-spectrogram.

\paragraph{Rectified flow transformers for music generation.}

Our architecture builds upon the MMDiT \cite{esser2024scaling} and Flux architecture. Specifically, we first construct an input sequence consisting of embedding of the text and noised music. The noised latent music representation $X_{spec} \in \mathbb{R}^{h\times w \times c}$ is flatten 2$\times$2 patches to a sequence of length $\frac{1}{2}h \cdot \frac{1}{2}w$. After aligning the dimensionality of the patch encoding and the fine-grained text encoding $c$, we concatenate the two sequences. 

We then forward with two type layers including double stream block and single stream blocks. In the double stream block, we employ two distinct sets of weights for the text and music modalities, effectively treating them as independent transformers that merge during the attention operation. This allows each modality to maintain its own space while still considering the other. 
In the single stream block, the text component is dropped, focusing solely on music sequence modeling with modulation.  
It is also found in AuraFlow\footnote{https://blog.fal.ai/auraflow/} that removing some of MMDiT layers to just be single DiT block were much more scalable and compute efficient way to train these models.

We incorporate embeddings of the timestep $t$ and coarse text $y$ into the modulation mechanism. 
Drawing from \cite{esser2024scaling}, we employ multiple text encoders to capture various levels of textual information, thereby enhancing overall model performance and increasing flexibility during inference. By applying individual dropout rates during training, our model allows the use of any subset of text encoders during inference. This flexibility extends to the ability to pre-store blank textual representations, bypassing the need for network computation during inference.

\subsection{Discussion}

\paragraph{Model at Scale.}
In summary, the hyper-parameters of proposed FluxMusic architecture include the following key elements:  
the number of double stream blocks $m$, number of single stream blocks $n$, hidden state dimension $d$, and attention head number $h$. 
Various configurations of FluxMusic are listed in Table \ref{tab:scale}, span a broad range of model sizes and computational requirements, from 142M to 2.1B parameters and from 194.5G to 2928.0G Flops. This range provides a thorough examination of the model's scalability.  
Additionally, the Gflop metric, evaluated for a  16$\times$128 text-to-music generation with a patch size of $p=2$, i.e., 10s music clips according to blank text, is calculated using the \texttt{thop} Python package.

\paragraph{Synthetic data incorporation.}

It is widely recognized that synthetically generated captions can greatly improve performance of generative model at scale, i.e., text-to-image generation \cite{fei2024dimba,chen2023pixart,betker2023improving,fei2021partially,fei2022deecap}.  
We follow their design and incorporate enriched music captions produced by a fine-tuned large language model.  
Specifically, we use the LP-MusicCaps model \cite{doh2024enriching,doh2023lp}, available on  Huggingface\footnote{https://huggingface.co/seungheondoh/ttmr-pp}.
To mitigate the potential risk of the text-to-music model forgetting certain concepts not covered in the music captioner’s knowledge base, we maintain a balanced input by using a mixture of 20\% original captions and 80\% synthetic captions.

\section{Experiments}

\subsection{Experimental settings}
\paragraph{Datasets.} 
We employ several datasets, including AudioSet Music Subset (ASM) \cite{gemmeke2017audio}, MagnaTagTune, Million Song Dataset (MSD) \cite{bertin2011million}, MagnaTagTune (MTT) \cite{law2009evaluation}, Free Music Archive (FMA) \cite{defferrard2016fma}, Music4All \cite{santana2020music4all}, and an additional private dataset. 
Each audio track was segmented into 10-second clips and uniformly sampled at 16 kHz to ensure consistency across the datasets. Detailed captions corresponding to these clips were sourced from Hugging Face datasets\footnote{https://huggingface.co/collections/seungheondoh/enriching-music-descriptions-661e9342edcea210d61e981d}. Additionally, we automatically labeled the remaining music data using LP-MusicCaps models. This preprocessing resulted in a comprehensive training dataset encompassing a total of $\sim$22K hours of diverse music content.

To benchmark our MusicFlux model against prior work, we conducted evaluations using the widely recognized MusicCaps dataset \citep{agostinelli2023musiclm} and the Song-Describer-Dataset \citep{manco2023song}. The prior dataset comprises 5.5K clips of 10.24 seconds each, accompanied by high-quality music descriptions provided by ten professional musicians. The latter dataset contains 706 licensed high-quality music recordings.

\paragraph{Implementail details.}
We utilize the last hidden state of FLAN-T5-XXL as fine-grained textual information and the pooler output of CLAP-L as coarse textual features. 
Referring to \cite{liu2024audioldm}, our training process involves 10-second music clips, randomly sampled from full tracks. The training configuration includes a batch size of 128, a gradient clipping threshold of 1.0, and a learning rate of 1e-4.
During inference, we apply a rectified flow with 50 steps and use a guidance scale of 3.5. To ensure model stability and performance, we maintain a secondary copy of the model weights, updated every 100 training batches through an exponential moving average (EMA) with a decay rate of 0.99, following the approach outlined by \citet{Peebles_2023}. For unconditional diffusion guidance, we independently set the outputs of each of the two text encoders to null with a probability of 10\%.

\paragraph{Evaluation metrics.}

The generated results are assessed using several objective metrics, including the Fréchet Audio Distance (FAD)~\citep{kilgour2018fr}, Kullback-Leibler Divergence (KL), Inception Score (IS). 
To ensure a standardized and consistent evaluation process, all metrics are calculated utilizing the \texttt{audioldm\_eval} library \cite{liu2024audioldm}.

\begin{figure}[t]
  \centering
   \includegraphics[width=0.98\linewidth]{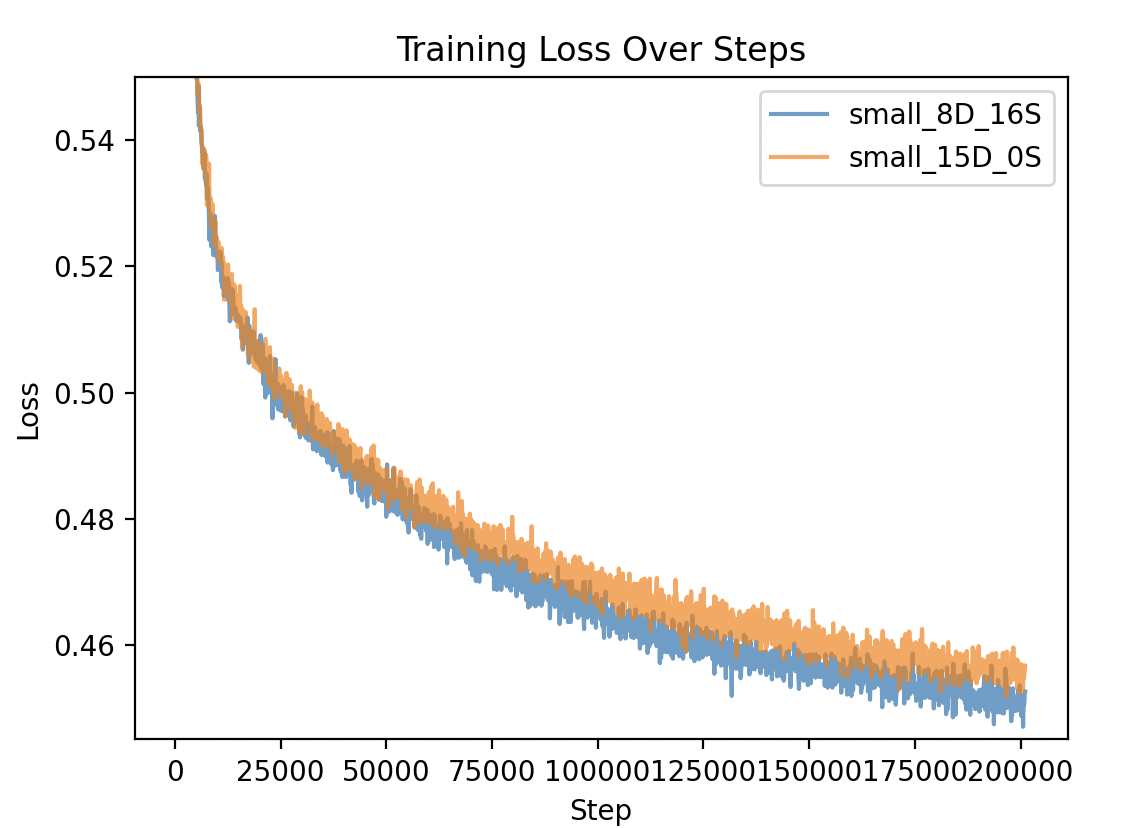}
   \caption{\textbf{The loss curve of different model structure with similar parameters.}  We can see that combine double and single stream block is much more scalable and compute efficient way for music generation model. 
   }
   \label{fig:stream} 
\end{figure}

\begin{figure}[t]
  \centering
   \includegraphics[width=0.98\linewidth]{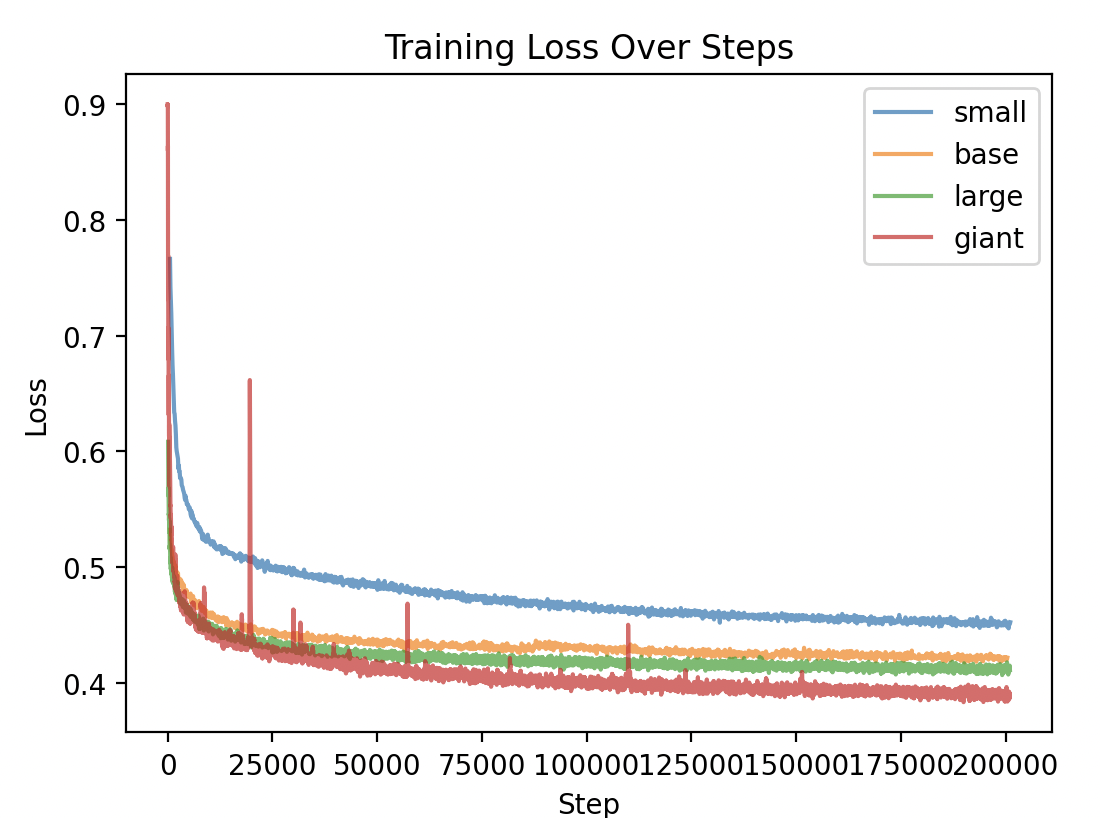}
   \caption{\textbf{The loss curve of different model parameters with same structure. } We can see that increase model parameters consistently improve the generative performance. 
   }
   \label{fig:scale} 
\end{figure}

\subsection{Model Analysis}

To assess the effectiveness of the proposed approaches and compare different strategies, we isolate and test only the specific components under consideration while keeping all other parts frozen. Ablation studies are performed on a subset of the training set, specifically utilizing the ASM and FMA datasets. For evaluation purposes, we employ an out-of-domain set comprising 1K samples randomly selected from the MTT dataset.

\begin{table}[t]
	\begin{center}
		\setlength{\tabcolsep}{4.mm}{
			\begin{tabular}{lccc}
			\toprule
 & FAD$\downarrow$   & IS$\uparrow$  & CLAP$\uparrow$ \\ \midrule
DDIM  & 7.42  & 1.67 & 0.201 \\
RF & 5.89  &2.43 & 0.312  \\
\bottomrule
			\end{tabular}}
	\end{center}
 {\caption{\textbf{Effect of rectified flow training in text-to-music generation.}  We train small version of FluxMusic with different sampling schedule and results show the superiority of RF training with comparable computation burden. }
				\label{tab:loss}}
\end{table}

\begin{table*}[ht] 
  \scalebox{1}{ 
  \begin{tabular}{lcccccccccc}
    \toprule
    & \multicolumn{2}{c}{\text{Details}} & \multicolumn{4}{c}{\text{MusicCaps}} & \multicolumn{4}{c}{\text{Song Describer Dataset}} \\
    \cmidrule(lr){2-3} \cmidrule(lr){4-7} \cmidrule(lr){8-11}
      & \text{Params} & \text{Hours} & \text{FAD} \(\downarrow\) & \text{KL} \(\downarrow\) & \text{IS} \(\uparrow\) & \text{CLAP} \(\uparrow\) & \text{FAD} \(\downarrow\) & \text{KL} \(\downarrow\) & \text{IS} \(\uparrow\) & \text{CLAP} \(\uparrow\) \\
    \midrule
    MusicLM & 1290M & 280k & 4.00 & - & - & - & - & - & - & - \\
    MusicGen  & 1.5B & 20k & 3.80 & 1.22 & - & 0.31 & 5.38 & 1.01 & 1.92 & 0.18 \\
    Mousai & 1042M & 2.5k & 7.50 & 1.59 & - & 0.23 & - & - & - & - \\
    Jen-1 & 746M & 5.0k & 2.0 & 1.29 & - & 0.33 & - & - & - & - \\
    AudioLDM 2 (Full) & 712M & 17.9k & 3.13 & {1.20} & - & - & - & - & - & - \\
    AudioLDM 2 (Music)  & 712M & 10.8k & 4.04 & 1.46 & 2.67 & 0.34 & 2.77 & 0.84 & 1.91 & 0.28 \\
    QA-MDT (U-Net) & 1.0B & 12.5k & 2.03 & 1.51 & 2.41 & 0.33 &  {1.01} & {0.83} & 1.92 & 0.30 \\
    QA-MDT  (DiT) & {675M} & 12.5k & {1.65} & 1.31 & {2.80} & {0.35} & 1.04 & {0.83} & {1.94} & {0.32} \\\midrule 
    FluxMusic & 2.1B &22K & 1.43 & 1.25 & 2.98 & 0.36 & 1.01 & 0.83 & 2.03 & 0.35\\
    \bottomrule
  \end{tabular}
}
 \centering
  \caption{\textbf{Evaluation results for text-to-music generation with diffusion-based models and language-based models.} We can see that with compettive parameters and training data, FluxMusic achieve best results in most metrics, demonstrating the promising of structure. }
  \label{tab:main} 
\end{table*}

\begin{table}[ht]
  \setlength{\tabcolsep}{8pt} 
  
  \begin{tabular}{lcccc}
    \toprule
    & \multicolumn{2}{c}{\text{Experts}} & \multicolumn{2}{c}{\text{Beginners}} \\
    \cmidrule(lr){2-3} \cmidrule(lr){4-5} 
    \text{Model} & \text{OVL} & \text{REL} & \text{OVL} & \text{REL} \\
    \midrule
    Ground Truth & 4.20 & 4.15 & 4.00 & 3.85 \\
    \midrule 
    AudioLDM 2 & 2.55 & 2.45 & 3.12 & 3.74  \\
    MusicGen & 3.13 & 3.34 & 3.06 & 3.70 \\
    FluxMusic & 3.35 & 3.54 & 3.25 & 3.80  \\
    \bottomrule
  \end{tabular}
  \centering
  \caption{\textbf{Evaluation results of text-to-music performances in human evaluation.} We denoted for text relevance (\text{REL}) and overall quality (\text{OVL}), with higher scores indicating better performance. }
  \label{tab:human}
\end{table}

\paragraph{Advantage of model architecture.}

We examine the architectural design choices for the diffusion network within FluxMusic, focusing on two specific variants: (1) utilizing double stream blocks exclusively throughout the entire network, and (2) employing a combination of both double and single stream blocks. In particular, we use a small version of the model, with one variant comprising 15 double stream blocks, referred to as 15D\_0S, and the other combining 8 double stream blocks with 16 single stream blocks, referred to as 8D\_16S. These configurations result in parameter counts of approximately 145.5M and 142.3M, respectively.
As depicted in Figure \ref{fig:stream}, the combined double and single modality stream block architecture not only accelerates the training process but also enhances generative performance, despite maintaining a comparable parameter scale. Consequently, we designate the mixed structure as the default configuration.

\paragraph{Effect of rectified flow.}

Table \ref{tab:loss} presents a comparative analysis of various training strategies employed in FluxMusic, including DDIM and rectified flow, using the small model version. Both strategy training with 128 batch size and 200K training steps to maintain an identical computation cost. As anticipated, and in line with prior research \cite{esser2024scaling}, rectified flow training demonstrates a positive impact on generative performance within the music domain.

\paragraph{Effect of model parameter scale.}

We examine the scaling properties of the FluxMusic framework by analyzing the impact of model depth, defined by the number of double and single stream layers, and model width, characterized by the hidden size dimension. Specifically, we train four variants of FluxMusic using 10-second clips, with model configurations ranging from small to giant, as detailed in Table \ref{tab:scale}.
As the loss cureve depicted in Figure \ref{fig:scale}, performance improves as the depth of double:single modality block increases from 8:16 to 16:32, and similarly, expanding the width from 512 to 1408 results in further performance gains. It is important to note that the model's performance has not yet fully converged, as training was conducted for only 200K steps. Nonetheless, across all configurations, substantial improvements are observed at all training stages as the depth and width of the FluxMusic architecture are increased.

\subsection{Compared with Previous Methods}
We conducted a comparative analysis of our proposed MusicFlux method against several prominent prior text-to-music approaches, including AudioLDM 2~\citep{liu2024audioldm}, Mousai~\citep{schneider2024mousai}, Jen-1~\citep{li2024jen}, and QA-MDT~\cite{li2024quality}, which model music using spectral latent spaces, as well as MusicLM~\citep{agostinelli2023musiclm} and MusicGen~\cite{copet2024simple}, which employ discrete representations. All the results of these comparisons are summarized in Table~\ref{tab:main}.

The experimental outcomes highlight the significant advantages of our FluxMusic models, which achieve state-of-the-art performance across multiple objective metrics. These findings underscore the scalability potential of the FluxMusic framework, particularly as model and dataset sizes consistently increase.
Although FluxMusic exhibited a slight advantage in FAD and KL metrics on the Song-Describer-Dataset, this may be attributed to instabilities stemming from the dataset's limited size. Further, our superiority in text-to-music generation was corroborated through additional subjective evaluations.

\begin{figure*}[t]
  \centering
   \includegraphics[width=0.96\linewidth]{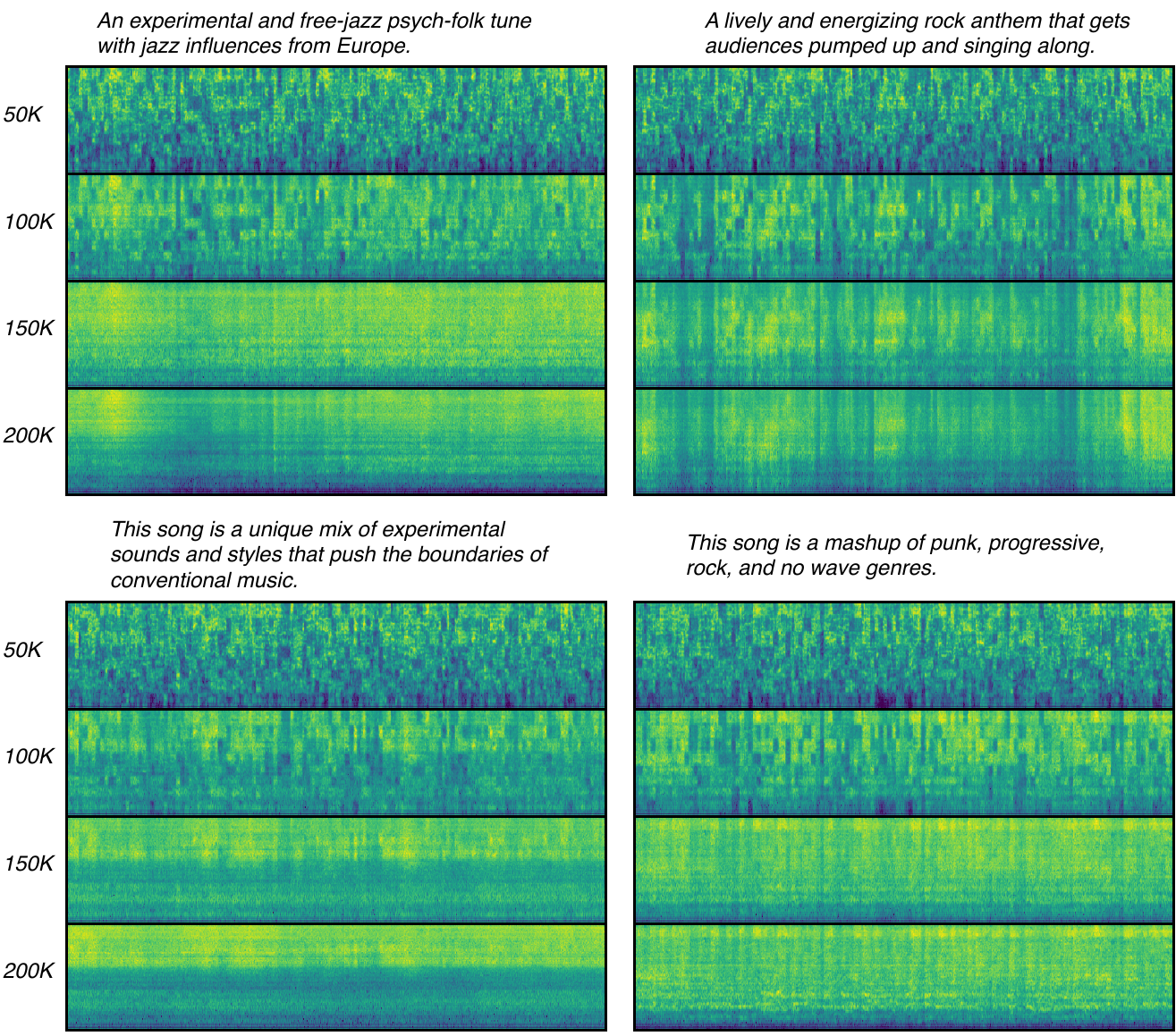}
   \caption{\textbf{Generated mel-spectrum cases of different training steps. } We plot small version of MusicFlux at every 50K training steps and we can find that the image becomes orderly and fine-grained instead of random and disorderly with the training continues. 
   }
   \label{fig:step} 
\end{figure*}

\begin{figure*}[t]
  \centering
   \includegraphics[width=0.96\linewidth]{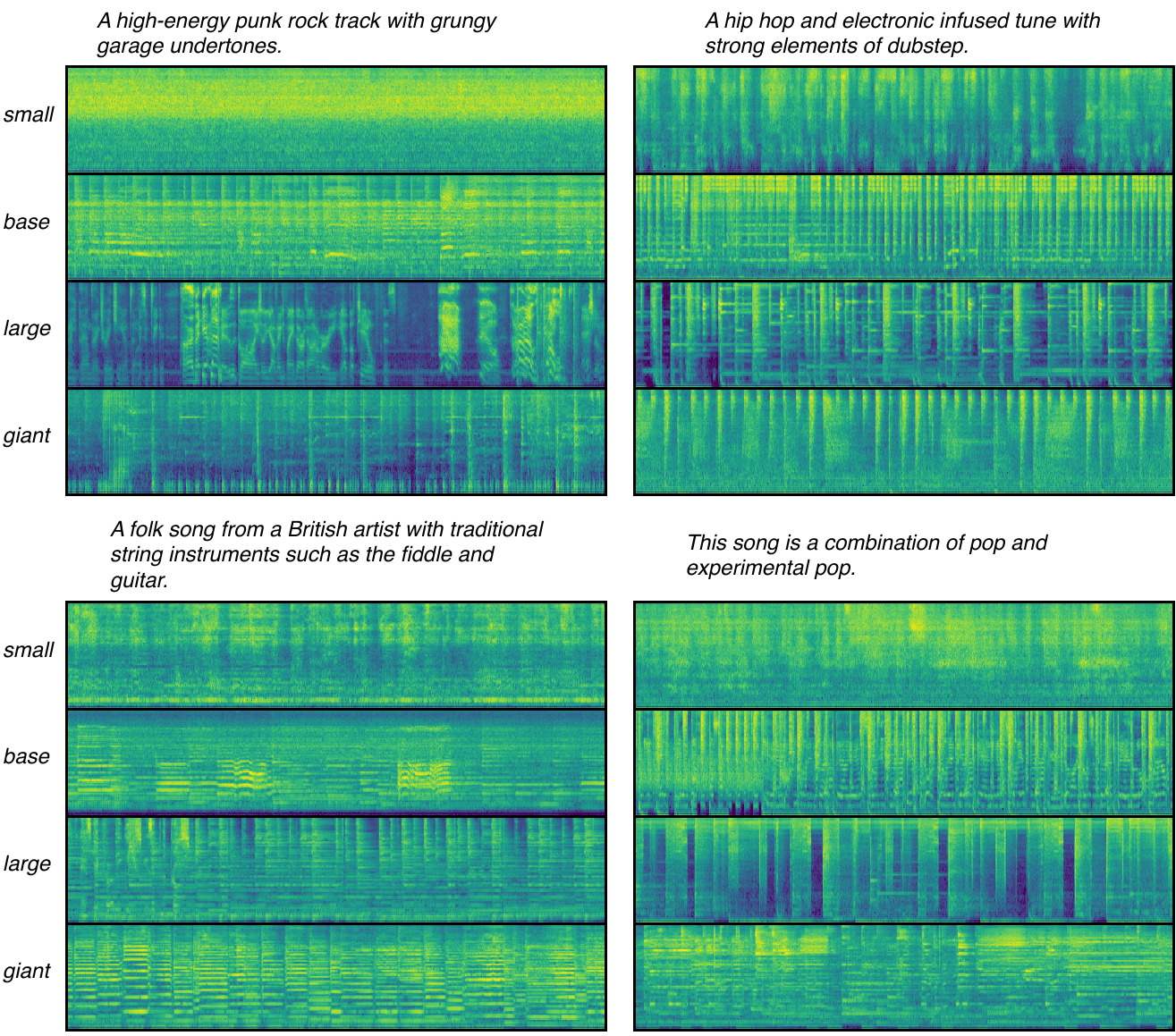}
   \caption{\textbf{Generated mel-spectrum cases of different model parameters. } With model size increase, the resulting mel-spectrum becomes more content rich and rhythmically distinct.
   }
   \label{fig:case_scale} 
\end{figure*}

\begin{figure*}[t]
  \centering
   \includegraphics[width=0.95\linewidth]{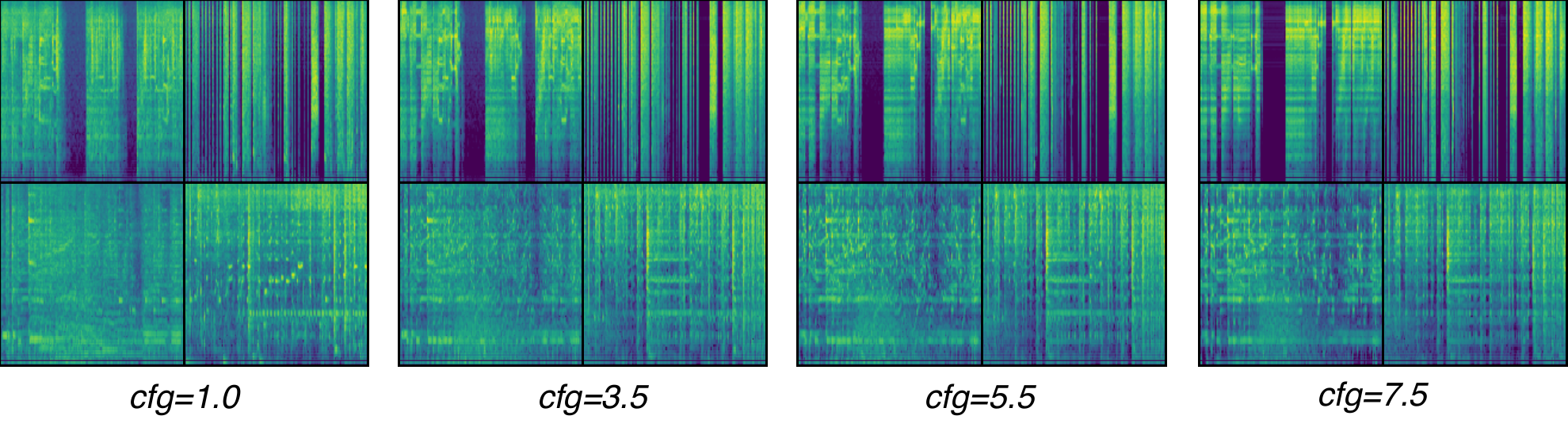}
   \caption{\textbf{Generated mel-spectrum cases of different classifier-free guidance.} We plot four clips with diverse textual prompts from small version of FluxMusic model and concatenate them in one figure. It can be seen that increasing the CFG number results in a more pronounced contrast in the generated mel-spectrum. To consistent with previous works, we set CFG=3.5 by default. 
   }
   \label{fig:cfg} 
\end{figure*}

\subsection{Human Evaluation}
We laso conducted a human evaluation, following settings outlined in \cite{li2024quality,liu2024audioldm}, to assess the performance of text-to-music generation. This evaluation focused on two key aspects of the generated audio samples: (i) overall quality (OVL) and (ii) relevance to the textual input (REL).
For the overall quality assessment, human raters were asked to evaluate the perceptual quality of the samples on a scale from 1 to 5. Similarly, the text relevance test required raters to score the alignment between the audio and the corresponding text input, also on a 1 to 5 scale.
Our evaluation team comprised individuals from diverse backgrounds, including professional music producers and novices with little to no prior knowledge in this domain. These groups are categorized as experts and beginners. Each randomly selected audio sample was evaluated by at least ten raters to ensure robust results.

As reflected in Table~\ref{tab:human}, our proposed FluxMusic method significantly enhances both the overall quality of the music and its alignment with the text input. These improvements can be attributed to the RF training strategy and the advanced architecture of our model. Notably, the feedback from experts indicates substantial gains, highlighting the model's potential utility for professionals in the audio industry.

\subsection{Visualization and Music Examples}

For a more convenient understanding, we visualize some generated music clips towards different prompt, from different perspective. These visualizations encompass: (1) different training step, (2) model parameter at scale, (2) setting of classifier-free guidance (CFG) number, the results are presented in Figure \ref{fig:step}, \ref{fig:case_scale}, and \ref{fig:cfg}, respectively. For more cases and listen intuitively, we recommand to visit the project webpage. 

\section{Conclusion}

In this paper, we explore an extension of the FLUX framework for text-to-music generation. Our model, FluxMusic, utilizes rectified flow transformers to predict mel-spectra iteratively within a latent VAE space. Experiments demonstrate the advanced performance comparable to existing benchmarks. Moreover, our study yields several notable findings: first, a simple rectified flow transformer performs effectively for audio spectrograms. Then, we identify the optimal strategy and learning approach through an ablation study. Future research will investigate scalability using a mixture-of-experts architecture and distillation techniques to enhance inference efficiency.


\bibliography{main}
\bibliographystyle{plainnat}

\end{document}